\documentclass[12pt]{iopart}

\usepackage{lineno,hyperref}
\usepackage{iopams}
\expandafter\let\csname equation*\endcsname\relax

\expandafter\let\csname endequation*\endcsname\relax

\usepackage{amsmath}

\usepackage{graphicx}  
\begin{document}
  
\title{Levitating frogs, machine learning and elliptic integrals.}

\author{M. Array\'as, J. L. Trueba and C. Uriarte}

\address{\'Area de Electromagnetismo, Universidad Rey Juan Carlos, Tulip\'
an s/n, 28933 M\'ostoles, Madrid, Spain}
\ead{manuel.arrayas@urjc.es}
\vspace{10pt}

\begin{abstract}
 We present the calculation of the stability region of a perfect diamagnet levitated in a magnetic field created by a circular current loop. We make use of the machine learning technique of automatic differentiation to illustrate the calculations with the elliptic integrals involved.   
\end{abstract}

\section{Introduction}
There is a growing interest in the so called data science. Modern computers and clever algorithms have made possible to deal with massive amount of data and extract information for making more and more sophisticated models and predictions. The magic words are ``deep learning'' or ``machine learning''. These are fancy words to denote a set of tools that comprise linear algebra, calculus, statistics and optimization. Employers demand deep learning experts so for physics students some basic knowledge would be of interest. Here we present a problem which serves to introduce the student in those techniques of scientific computing while keeping the interest in physics topics. It is inspired on the celebrated paper ``Of flying frogs and levitrons'' \cite{Berry285} which has been the guiding idea for the design of a levitating device to explore superfluity regimes \cite{Arr21}. 

In order to levitate diamagnetic materials, it is crucial to find the stability region of a given magnetic field configuration. The problem reduces to identify the place where the energy of the system obtains a minimum. To achieve levitation the total force on the object must vanish thus the gradient of the energy should be zero. However that condition is not enough, we also need to impose that the Laplacian of the energy will be positive, so the equilibrium is stable. Solving the problem involves to get the magnetic fields, and their derivatives, mostly in forms of gradients and Hessians. Here is where some techniques of machine learning are applied, in particular automatic differentiation (AD) which is at the heart of deep learning \cite{Baydin18}. 

We will study the particular case of finding the stability region in the field created by a circular current loop. The problem presented here gives us an extra bonus: the elliptic functions. There are available ready-to-try packages (both free and commercial) for deep learning applications ranging from image and speech recognition, genome-based taxonomic classification, drug research and so on. Normally the standard functions used in these packages are limited to a set of basic functions, like the sigmoids \cite{Higham19}. When considering the magnetic field created by a circular loop we will be dealing with elliptic functions. We will code explicitly the AD rules for elliptic integrals of the first and second kind, and that brings us an excellent opportunity to introduce the fascinating subject of elliptic functions to the student \cite{Lawden89}.

\section{The field of a circular current loop}
First let us summarize the classical solution of finding the field created by a circular current loop of radius $a$ \cite{Jackson98}. We take our origin of coordinates at the center of the loop and $xy$ the plane containing the loop.  Thus the current density is given by $Id{\bf s}= -I\sin{\phi}{\bf i}+I\cos{\phi}{\bf j}$, the position of the loop by $ {\bf r}_i=a\cos{\phi}{\bf i}+a\sin{\phi}{\bf j}$, and defining  $R=|{\bf r}-{\bf r}_i|$, we have  \[ {\bf A( {\bf r})}=\frac{\mu_0 I}{4\pi}\oint\frac{d{\bf s}}{R}, \] for the vector potential in the Coulomb gauge. Taking cylindrical coordinates $(r,\phi,z)$, it is an exercise to prove \cite{Landau} that

\begin{equation}
     \begin{split}
       A_\phi&= \frac{\mu_0}{4\pi} \frac{4aI}{\sqrt{(r+a)^2+z^2}} \left(\frac{2-k^2}{k^2}K(k^2)-\frac{2}{k^2}E(k^2) \right)\\
       &=  \frac{\mu_0}{4\pi}{4I}\sqrt{\frac{a}{r}} \left[\left(\frac{1}{k}-\frac{k}{2}\right)K(k^2)-\frac{1}{k}E(k^2) \right],
     \end{split}
     \label{potential}
     \end{equation}
where $k^2=4ar/[(r+a)^2 + z^2]$ and it is defined the elliptic integrals of the first and second type as
     \begin{equation}
       K(m)=\int_0^{\frac{\pi}{2}} \frac{d\theta}{\sqrt{1-m\sin^2{\theta}}}, \,\,\, E(m)=\int_0^{\frac{\pi}{2}}\sqrt{1-m\sin^2{\theta}}\, d\theta,
       \label{elliptic}
     \end{equation}
The magnetic field can be calculated as
     \begin{equation}
       B_r=-\frac{\partial A_\phi}{\partial z},\,\,\, B_z=\frac{1}{r}\frac{\partial (r A_\phi)}{\partial r},
     \label{field}
     \end{equation}
which already involves derivatives of the elliptic integrals. 
     
\section {The stability conditions for a perfect diamagnet}
Having obtained the magnetic field we introduce the stability conditions for a perfect diamagnet \cite{Berry285}. When a isotropic linear magnetic material of mass $m$ is introduced in a magnetic field, the induced magnetic moment reads
\begin{equation}
  \label{eq:moment}
  {\bf m}({\bf r})= \frac{\chi {\bf B}({\bf r})}{\mu_0},
\end{equation}
and the change in the free energy of the system kept at constant temperature reads $d{f}=mgz + {\bf m}\cdot d{\bf B}$. For a perfect diamagnet, we have $\chi=-1$ and the free energy reads 
\begin{equation}
  f(T,B)=f_0 + mgz - \frac{B^2}{2\mu_0},
  \label{free}
\end{equation}
where $f_0$ is the free energy of the body at zero fields and $B^2$ is the squared modulus of the magnetic field. At equilibrium we must have a minimum, so the conditions to fulfill are $d{f} = 0$ and $d^2{f}<0$. The second condition provides the stability of the levitation and using \eqref{free} turns into
\begin{equation}
  \nabla^2 B^2 > 0.
  \label{stability}
\end{equation}

\section{Automatic differentiation (AD) and deep learning}
Computing the stability condition involves several times differentiation. There are few techniques to calculate derivatives with a computer. First we can resort to manually working out the derivatives and coding them. Alternatively we can use a computer algebra system, free (Maxima or Sympy) or commercial (Maple or Mathematica) to make the symbolic manipulations. We also can use finite difference approximations to get the numerical derivatives. Finally there is the technique of automatic differentiation (AD) to perform the calculations.

Numerical differentiation based on finite difference approximations are easy to implement but can be inaccurate due to truncation errors. Normally for computing gradients and Hessian one can use an interpolation scheme, but can be highly inefficient and the scaling is very poor \cite{Baydin18}. Manual differentiation and symbolic methods often result in rather long expressions as the complexity of derivatives computed by symbolic differentiation grows exponentially \cite{Corliss88} and afterwards we need to evaluate numerically the final expression. To address those issues AD was invented as it allows the evaluation of derivatives at machine precision. The importance of that in deep learning and neural nets comes from the following. In deep learning we want a function that classifies training data correctly, so we can use to make predictions based on new data \cite{Strang19}. If we arrange the inputs in a vector ${\bf x}$ and the outputs in a vector ${\bf y}$ we get the deep learning function ${\bf y}=F({\bf x})$. Normally the deep learning function have the form $F({\bf x})=L_k(N_{k-1}(L_{k-1}(N_{k-1}(...N_1(L_1({\bf x}))))))$ where $L$ and $N$ are linear and nonlinear functions. The linear functions $L$ are affine functions $L_i({\bf x}) = A_i {\bf x} +  {\bf b}_i$ where the $A_i$ and ${\bf b}_i$ are called the weights, and $i$ runs across the so called neural layers. The problem in deep learning is to find the weights to minimize a total loss function over the sample data \cite{Strang19}. This is  an optimization problem like finding the stability region for the levitation of our diamagnet. 

In AD there are two variants. The forward mode differentiation, represented symbolically as $(\partial/\partial X)$ and the reverse mode or back propagation represented as $(\partial Z/\partial)$ \cite{Olah15}. AD is based on the basic fact that ultimately all numerical computations involve a finite set of elementary operations. For example to compute $E(k^2)$ defined in section 2 the finite steps needed are
\begin{equation}
  \begin{split}
y_1 &= 4a\\
y_2 &= y_1 * r\\
y_3 &= r + a\\
y_4 &= y_3^2\\ 
y_5 &= y_2 / y_4\\
y_6 &= z^2\\
y_7 &= y_5 + y_6\\
y_8 &= E(y_7)
\end{split}
\end{equation}
Those steps constitute a Wengert list \cite{Wengert64} which can be represented in a directed computational graph. When calculating derivatives, the graph is augmented with the extra calculations involved. The computational problem is how to travel the graph in an efficient way and that determines the two implementations of AD, the forward and the backward mode \cite{Baydin18}.   

\section{Dual numbers and the rules for the elliptic integrals}
Forward mode can be implemented very efficiently using dual numbers \cite{milk}. When we calculate the derivatives, the number of terms in the Wengert list starts growing. For example, compare the lists to compute $E(x^2)$,
\begin{equation}
  \begin{split}
y_1 &= x ^ 2 \\
y_2 &= E(y_1)
  \end{split}
  \label{a}
\end{equation}
and $dE(x^2)/dx$,
\begin{equation}
  \begin{split}
y_1 &= x ^ 2 \\
y_2 &= E(y_1) \\
dy_1 &= x ^ 1 \\ 
dy_2 &= 2 * dy_1 \\
dy_3 &= 2 * y_1\\
dy_4 &= y_2 / dy_3\\
dy_5 &= dy_4 * dy_2\\
dy_6 &= K(y_1)\\
dy_7 &= -dy_6\\
dy_8 &= 2 * y_1\\
dy_{9} &= dy_7 / dy_8\\
dy_{10} &= dy_{9} * dy_1\\
dy_{11} &= dy_5 + dy_{10}
  \end{split}
\end{equation}

We could try to remove from the list the values that will not be needed in further steps, for example after evaluating $dy_5$ we could forget about $dy_4$. However, $y_2$ is used at step $dy_4$ so it needs to be kept during further steps. Dual numbers provide evaluation of Wengert list in a very efficient way.   

A dual number is like a complex number, where the imaginary symbol $i$ is replaced by another symbol $\epsilon$ being a nilpotent number of index 2, i.e $\epsilon^2 =0$. So we can write any dual number as $a+b\epsilon$. If for any function $g(x)$ we take the truncated Taylor series expansion \cite{Seppo76} and write it as $g + g'\epsilon$ this dual representation will propagate the usual rules of differentiation, for example the product rule
\[(g + g'\epsilon)(h + g'\epsilon) = gh + (gh'+hg')\epsilon.\]
Now let us extend any real function $g(x)$ to dual numbers by defining \[g(a+b\epsilon)\equiv g(a) + g'(a)b\epsilon.\] The chain rule will propagate automatically using the algebra of dual numbers when taking as the argument the dual number $x+\epsilon$,
\[g(h(x+\epsilon))=g(h(x)+h'(x)\epsilon) = g(h(x))+g'(h(x))h'(x)\epsilon.\]
Thus to implement the extension of the real functions to dual numbers we have to provide explicitly the derivative of the functions involved. In the case of elliptic integrals, from \eqref{elliptic} it is straightforward to obtain
\begin{eqnarray}
  K'(m)=\frac{dK(m)}{dm}=\frac{E(m)}{2m(1-m)}-\frac{K(m)}{2m},\\
  E'(m)=\frac{dE(m)}{dm}=\frac{E(m)-K(m)}{2m}. 
  \label{elliptic_deriv}
\end{eqnarray}

For instance to find the magnetic field created by the circular loop \cite{Landau} we compute the derivative of $E(k^2)$ following the dual number rules,
\[E((x+\epsilon)^2)=E(x^2+2x\epsilon)=E(x^2)+2xE'(x^2)\epsilon,\]
and substituting \eqref{elliptic_deriv} 
\[\frac{dE(k^2)}{dk}=\frac{E(k^2)-K(k^2)}{k}.\]
In the same way the derivative of $K(k^2)$ is calculated,
\[\frac{dK(k^2)}{dk}=\frac{E(k^2)}{k(1-k^2)}-\frac{K(k^2)}{k}.\]
Those expressions allow to find the explicit expression of the magnetic field using \eqref{field} and \eqref{potential}.

\section{Computing the stability regions for the circular current loop configuration}
We are ready to find the stability region of a perfect diamagnet in the magnetic field produced by a circular current loop. We take a cross section, calculate the left hand side of \eqref{stability} and code in grey colour the condition of being positive. We do that using the AD forward mode algorithm discussed in the previous section.

\begin{figure}
  \begin{centering}
\includegraphics[width=0.75\linewidth]{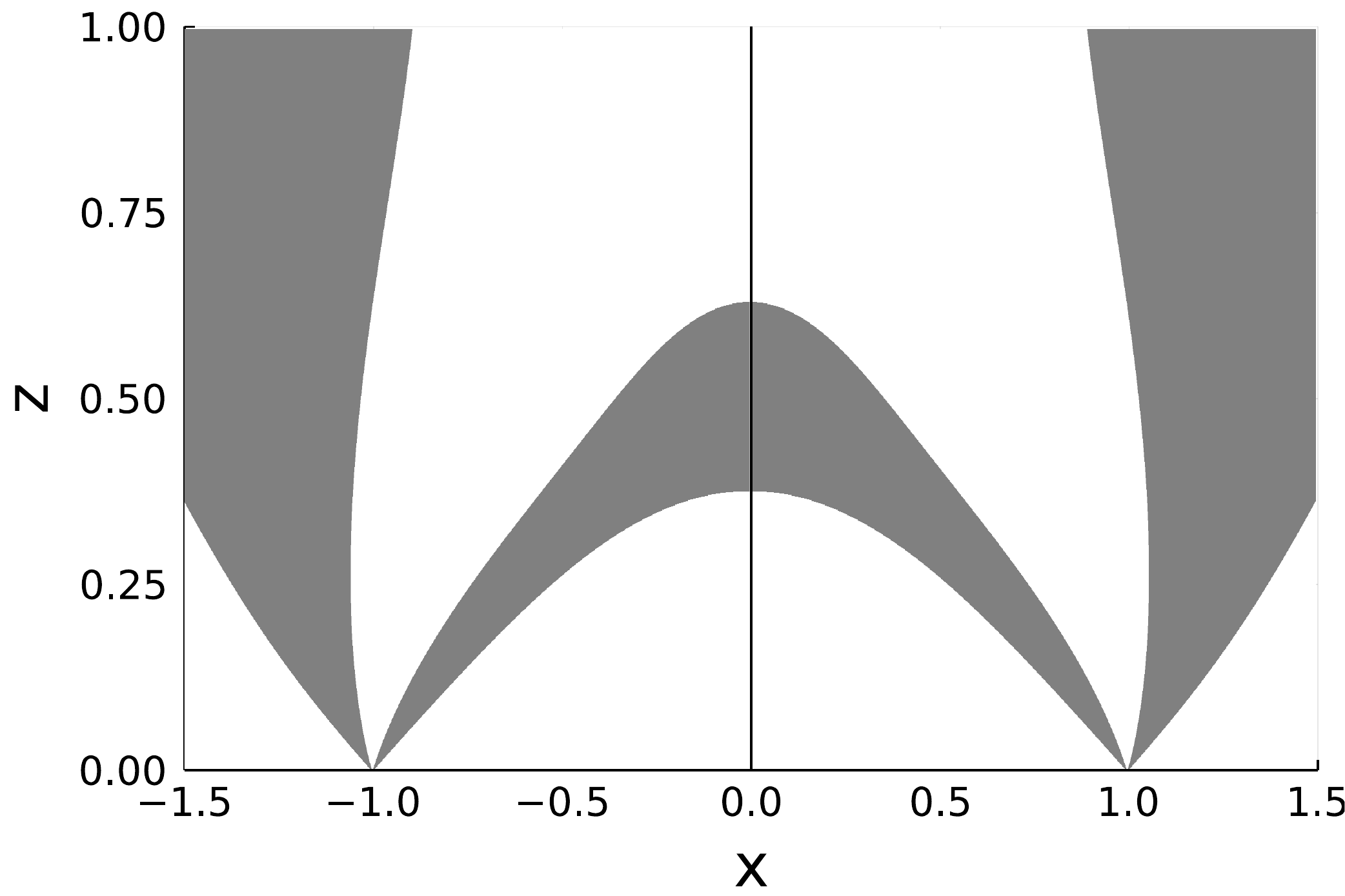}%
\caption{\label{stbregion}The transversal plane cut $XZ$. The shaded area is the stability region where $\nabla^2 B^2 > 0$.}
\end{centering}
  \end{figure}

The result is shown in Fig.~\ref{stbregion} where the $XZ$ plane is taken. The loop is in the $XY$ plane and cuts the $XZ$ plane at points $X=\pm 1$. In order to achieve stable levitation, the magnetic force must balance the gravity force in the shaded area depicted in the figure. That can be done by fine tuning the mass of the diamagnet and the current intensity circulating in the loop. 

A note on the code we wrote for this work. For the implementation of the dual number algorithm of the AD forward mode, we have made use of the Julia lenguage \cite{Julia}. Julia was designed from the beginning for high performance being many of its features well suited for numerical analysis. We recommend it as a pedagogical tool now being used in several universities and online courses. Moreover, it is free.

In particular we took the library (called package in Julia's argot) ``ForwardDiff.jl'' \cite{lib1}. That library depends on another library called ``DiffRules.jl'' \cite{lib2}. This last library contains the file ``rules.jl'', which implements the extension of the elementary functions to dual numbers. At present it lacks of the elliptic functions support, but it can be easily edited and added the expressions \eqref{elliptic_deriv}.     

\section{Discussion}
Inspired in the levitation of a frog \cite{Berry285} we have presented the problem of finding the region of stability of a perfect diamagnet in a magnetic field created by a circular current loop. This is a simplified version of the problem solved for the design of an actual levitation system of a superconductor immersed in a superfluid \cite{Arr21}.

The levitating problem offers an excellent opportunity to introduce the optimization methods encountered in deep learning such as computing gradients and Hessian and learn about this important topic. We have introduced the dual numbers and made use of them in order to implement the forward mode of the automatic differentiation technique.

Last but not least we hope that the case presented can bring some attention to the topic of elliptic functions. We just touched that point marginally considering the derivatives of the complete elliptic integral of the first and second types, but there is a treasure to discover in elliptic functions and elliptic curves \cite{Lawden89}.


\end{document}